
%
%
%
%
%

\input phyzzm

\theory

\REF\Kino{T.~Kinoshita and W.~B.~ Lindquist, \journal Phys. Rev.&D39,
 2407 (89); \journal&D42, 636 (90).}

\REF\Remi{M.~Caffo, E.~Remiddi and S.~Turrini,
 \journal Nucl. Phys.&B141, 302 (78); \journal Nuov. Cim.&79A, 220
 (84); A.~Hill, F.~Ortolani and E.~Remiddi, ``The Bound State Problem in
 QED,'' in {\sl The Hydrogen Atom}, G.~F.~Bassani, M.~Inguscio and
 {T.~Haen\nobreak%
 sch}, Eds., Springer Verlag, 1989.}

\REF\Stru{H.~Strubbe, \journal Comp. Phys. Commun.&8, 1 (74);
 \journal&18, 1 (79).}

%
%
\nopagenumbers
\theory
\pubtype{}  
\Pubnum={UM--TH--91--18}
\date={June 9, 1993}
\bigskip
\titlepage                                                         %
%
    \title{\bf Schoonschip '91}
    \author{Martinus J. G. Veltman and David N. Williams}
    \address{Randall Laboratory of Physics \nextline
             The University of Michigan \nextline
             Ann Arbor, MI 48109-1120}
\medskip
\abstract

The symbolic manipulation program Schoonschip is being made freely
available for a number of computers with Motorola 680x0 cpu's.  It can run
on machines with relatively modest memory and disk resources, and is
designed to run as fast as possible given the host constraints.  Memory and
disk utilization can be adapted to tune performance.  Recently added
capabilities include a system for efficient generation of diagrams, gamma
algebra for continuous dimensions, algorithmic improvements for handling
large problems, and an increase in the allowed number of {\tt X}
expressions.

\endpage %

%
\pagenumber=2
\pagenumbers

\chapter{History}

        The first working version of Schoonschip ran on an IBM 7094
in December, 1963.  A basic problem in particle physics, namely,
radiative corrections to photons interacting with a charged vector boson,
was successfully completed.  In 1966 Schoonschip was ported to the CDC 6600
series computers; and gradually, as hard disks and file storage became
available, the distribution extended to encompass virtually all existing
CDC machines.  Even quite recently this version was used for a very
complicated
calculation of higher order effects in quantum chromodynamics in the USSR.
Notable applications in quantum electrodynamics include works by
Kinoshita and coauthors,\refmark{\Kino} and Remiddi and
coauthors,\refmark{\Remi} among which we cite only a few.  The early
version of the Schoonschip manual by H.~Strubbe\refmark{\Stru} has
been the standard reference for citations of Schoonschip itself.

        In 1983, when microprocessors of sufficient capability became
available, Schoonschip was ported to a computer built around the Motorola
68000 microprocessor, the Charles River Data Systems machine.  The
operating system was UNIX-like, and the port to any other machine
containing a 68000 series processor and featuring a UNIX-type operating
system became trivial.


\chapter{Features}

        It is not likely that Schoonschip will ever have the popular
constituency of programs like Maple and Mathematica.  It does not have a
graphics interface, and has much less built-in.\foot%
{There does exist the beginning of a mathematics and particle
physics knowledge base of Schoonschip procedures, some of which is
provided in the current distribution.}%
It can output formulas in Fortran format, but not in TeX or PostScript
format.  Although there is no reason why it cannot be used for general
purposes, with a few exceptions its primary role so far has been that of a
scientific instrument (especially in particle physics); and it demands a
certain level of mathematical expertise for its effective use.

        In this section, we review some of the features that are
special to Schoonschip, or relatively new to it.

\section{Speed}

        Schoonschip is written entirely in assembly language, which not
only allows it to run fast, but also reduces its size compared to equivalent
programs compiled in a higher level language.

\section{Precision}

        Unlike some other symbolic manipulation programs, Schoonschip does
not use ``infinite'' precision arithmetic.  Its standard number is
represented in floating point, with about 29 decimal digits for the
mantissa and 4 decimal digits for the exponent.  The precision for
cancellation may be set by the user, with a default of about 25 decimal
digits.  The default output is in rational number format, within the
current precision.

        This approach has evolved after many years of trying different
schemes, and seems about optimum for most scientific applications.

\section{Noncommutative algebra}

         Schoonschip is efficient at noncommutative algebra, because
one of its major data types, the function, is noncommutative by default.

\section{Gamma algebra}

        The known gamma algebra algorithms are built-in, and provide for
efficient and compact evaluation.  The algorithms in recent versions have
been extended to continuous dimensions.

\section{Output format}

        The output format is fairly rigid, and not particularly adapted for
direct use in scientific publications.  The present version does, however,
have a few new options which offer more control over the output format for
formulas.

\section{Substitution efficiency}

        Schoonschip pioneered the substitution concept in algebraic
manipulation programs, and its speed and memory efficient approach is
still not common;\foot
{As far as we know, Form, by J.~Vermaseren, is the only other program that
currently follows a similar approach.}
indeed it seems incompatible with the Lisp heritage of several other
programs.

\section{Large problems}

        The handling of large problems has always been a major concern in
the design of Schoonschip, and recent versions have incorporated new
sorting algorithms and tuning facilities that offer order of magnitude
improvements in speed and disk utilization.  The number of {\tt X}
expressions allowed has been increased to about $1,000$ in the current
release.

\section{Conditionals}

        Recent versions of Schoonschip have a more familiar and flexible
treatment of conditional statements, including

        {\tt IF ... ELSE ... ENDIF }

\noindent constructs.

\section{Compatibility with earlier versions}

        Programs that worked with older M68000 and CDC versions of
Schoonschip will run under Schoonschip '91 after minor, mainly cosmetic
modifications, which are described in the manuals.

\section{Compatibility among machines}

        All versions run alike on the various machines mentioned in the
next section, except for differences in clock speed and host system
efficiency; and Schoonschip programs developed on one machine will run on
any of the others without change, beyond the usual translation of text
file formats.  Schoonschip programs are prepared with a normal text editor.

\section{Documentation}

        One of the text files in the free distribution is a comprehensive
manual, which is keyed to a set of program examples that is also included.
Apart from the free distribution, the authors of this article are
publishing a reorganized, indexed, hardcopy version of the manual.

\section{Examples}

        Nothing is more instructive than a working example.  Therefore a
large collection of programs, collected over the lifetime of Schoonschip,
has been put together, notably including the very first program, executed
in December, 1963.


\chapter{Machine Requirements}

        At present, Schoonschip binary executable files are available
for the following computers:  Amiga, Atari, NeXT, and Sun 3.  The
Motorola 680x0 microprocessor is common to these machines.\foot
{Schoonschip does not run, for example, on a Sun SparcStation or an IBM
PC.}
It runs on the 68000, 68020, 68030, and 68040 microprocessors, and does not
require a floating point coprocessor.  The standard memory requirement is
1~Megabyte of contiguous RAM workspace, plus a little over 110~Kilobytes for
the program.\foot
{The executable can be assembled to load with less workspace.  Problems of
moderate size run comfortably with less than 400K.}
Input and output file space is of course also needed.  Typically, a rather
large program might produce an output file of 400K, with temporary files
totaling 1~Megabyte.

        All of the machines listed below provide acceptable performance for
real problems.  At present, the NeXTstation, with its 25 MHz, 68040 cpu
gives the best performance by a substantial margin.  Second in performance
are the machines with 25 MHz, 68030 cpu's, such as the Amiga~3000 and
Mac~II.  As the 68040 cpu and higher clock speeds migrate to more
machines, the situation will of course change.  Very good performance has
become relatively inexpensive.

\section{Amiga}

        The Amiga version has been tested on an Amiga 2000 (68000 cpu)
with AmigaDOS 1.3, and on an Amiga 3000 (68030 cpu) with AmigaDOS
1.3 and 2.0.  There is no reason to expect it to fail on an Amiga 500 or
1000, but it has not been tested on those machines.

\section{Atari}

        The Atari version runs on an Atari ST1020 or Atari Mega.  It will
presumably run also on the new machine, containing a 68030 cpu and
coprocessor, but that has not been tested.

\section{Next}

        The NeXT version has been tested on a NeXTstation with a 68040
cpu and release 2.1 of the operating system.

\section{Sun}

        The Sun version has been tested on a Sun 3/140 with a 68020 cpu
and Sun UNIX~4.4 Release~3.3.

\section{Macintosh}

        A working Macintosh II version exists, but it has not yet been
upgraded to System 7.  It replaces the Macintosh graphics interface with a
complete CLI ({\it command line interface}) environment, which involves
rather more than the other ports.  A simpler version operating with the
standard graphics interface will be released in the near future.

\chapter{Availability}

        The Information Technology Division (ITD) of the University of
Michigan is providing (for the time being at least) an anonymous ftp host
for Schoonschip software, programs, and documentation, as well as other
physics related computational material.  Users with access to the internet
can copy these files to their host machines by logging on with the ftp
program available at many UNIX and VAX installations.  The internet address
is

        {\tt archive.umich.edu}

\noindent The software is currently in the directory {\tt physics/schip}.

        The transcript of an actual session run from a Sun follows, with
a little editing:

{\singlespace \obeylines \obeyspaces \tt
\noindent > ftp archive.umich.edu
Connected to archive.umich.edu.
220 earth.rs.itd.umich.edu FTP server
\indent  (Version 4.82 Wed Aug 14 23:53:47 EDT 1991) ready.
Name (archive.umich.edu:williams): ftp
331 Guest login ok, send ident as password.
Password:
230 Guest login ok, access restrictions apply.

\noindent ftp> cd physics/schip
250 CWD command successful.

\noindent ftp> dir
200 PORT command successful.
150 Opening data connection for /bin/ls (141.211.96.20,1274) (0 bytes).
total 1212
-rw-rw-rw-  1 9030     staff        1592 Aug  7 15:05 AmiAta.txt
-rw-rw-rw-  1 9030     staff        1313 Aug  7 15:05 INDEX
-rw-r--r--  1 9030     staff         587 Jul 17 19:24 Mac.txt
-rw-r--r--  1 9030     staff         600 Jul 17 19:24 NextSun.txt
-rw-rw-rw-  1 9030     staff        4608 Aug  5 15:03 README
-rw-rw-rw-  1 9030     staff        2746 Aug  5 15:04 README.FTP
-rw-rw-rw-  1 9030     staff      222076 Aug  6 13:32 SDocsAMI.LZH
-rw-rw-rw-  1 9030     staff      226017 Aug  6 13:50 SDocsATA.LZH
-rw-rw-rw-  1 9030     staff      100692 Aug  7 15:05 SchipAMI\_M.LZH
-rw-rw-rw-  1 9030     staff      100691 Aug  7 15:05 SchipAMI\_S.LZH
-rw-rw-rw-  1 9030     staff       82293 Aug  6 13:51 SchipATA.LZH
-rw-rw-rw-  1 9030     staff      287582 Aug  6 14:17 SchipDocs.tar.Z
-rw-rw-rw-  1 9030     staff      101291 Aug  6 14:12 SchipNXT.tar.Z
-rw-rw-rw-  1 9030     staff      103313 Aug  6 14:07 SchipSUN.tar.Z
226 Transfer complete.
952 bytes received in 1.4 seconds (0.65 Kbytes/s)

\noindent ftp> binary
200 Type set to I.

\noindent ftp> recv SchipATA.LZH
200 PORT command successful.
150 Opening data connection for SchipATA.LZH
\indent (141.211.96.20,1275) (82293 bytes).
226 Transfer complete.
local: SchipATA.LZH remote: SchipATA.LZH
82293 bytes received in 1.9 seconds (41 Kbytes/s)

\noindent ftp> quit
221 Goodbye.}

        The password supplied by the user is blanked out above.  It can
be essentially anything.

        A few of the files in the above list are uncompressed text, and
can be copied with the ftp {\tt recv} command without setting the ftp
binary (image) mode, namely, the files with a {\tt .txt} extension, and
the files {\tt README} and {\tt INDEX}.  These files describe what the
other files contain, and explain how to copy and decompress them.

        The software itself, the manual, and a number of sample Schoonschip
programs are in compressed formats, and must be copied in ftp binary mode,
as in the session above.

        Although copyrighted by M.J.V., the Schoonschip files at
{\tt archive.umich.edu} may be freely copied and redistributed, on a
not-for-profit basis.  Technically, they are ``freeware.''

        Amiga and Atari users whose computers are not networked to the
internet still have the problem of downloading the files after retrieval
to a local machine which does have internet access.  That is a common
problem, with a number of solutions, among which Kermit over the telephone
is ubiquitous although not entirely painless.  An entirely acceptable
solution is to get floppies from somebody who already has a copy.



\chapter{Support}

        Every effort will be made to keep the program up to date, but
we are not in a position to offer systematic support.  A lot of examples
have been provided, as well as some actual calculations.  We invite users
willing to make their Schoonschip calculations freely available to submit
programs to the anonymous ftp host {\tt archive.umich.edu}, which can
serve as a central repository.

        Problem reports and questions about availability of the software
and documentation may be directed to either of us, and especially to D.N.W.
at

        {\tt \singlespace
        David.N.Williams@um.cc.umich.edu\nextline
        \indent DWilliams@UMiPhys.bitnet\nextline
        \indent 7506.3124@CompuServe.com}

\endpage

\refout
\end